\documentclass[10pt,a4paper]{article}

\usepackage{tikz}
\usetikzlibrary{decorations.pathmorphing,calc}
\usetikzlibrary{positioning}
\usepackage{jheppub_kim}
\usepackage{pdflscape}
\usepackage{amsmath}
\usepackage{amssymb}
\usepackage{dcolumn}
\usepackage{bm}
\usepackage{color}
\usepackage{amsmath}
\usepackage{breqn}
\usepackage{graphicx}
\usepackage{epsfig}
\usepackage{caption,subcaption}
\usepackage{amsfonts}
\usepackage{graphicx}
\usepackage{parskip}
\usepackage{dcolumn}
\UseRawInputEncoding

\begin{document}

\title{Gravitational Collapse of Bose-Einstein condensate dark matter in Generalized Vaidya spacetime}

\author{Prabir Rudra}

\affiliation{Department of Mathematics, Asutosh College,
Kolkata-700 026, India}

\emailAdd{prudra.math@gmail.com}

\abstract{In this work we study the gravitational collapse procedure in generalized Vaidya spacetime with Bose-Einstein condensate dark matter density profile. We use the generalized Vaidya metric to simulate the spacetime of a big star and subsequently obtain the field equations. Then we proceed to determine the star system's mass parameter by solving the field equations. Then the gravitational collapse mechanism is investigated using the derived solutions. Investigating the nature of the singularity (if formed) as the end state of the collapse is the main goal. Dark matter in the form of Bose-Einstein condensate is expected to play a crucial role in the fate of the collapse. We see that there is a possibility of the formation of both black holes and naked singularities as the end state of the collapse depending upon the initial conditions. The junction conditions are derived with a Vaidya exterior and a Friedmann interior and some important insights are obtained. A Penrose diagram showing the causal relations between the spacetimes is generated and studied in detail.}

\maketitle

\section{Introduction}
In terms of understanding gravitational phenomena at both cosmological and astrophysical scales, Einstein's general relativity (GR) theory has proven to be remarkably successful. This theory of gravity makes significant advancements through the intrinsic gravitational systems, which include star and galaxy structures and account for a large fraction of the observable Universe. Understanding the development of these self-gravitating systems is essential to understanding the universe's age, composition, and other unknown details. One of the crucial phases in the evolution of celestial structures is the gravitational collapse process that leads to a celestial structure's demise and the creation of compact star objects. Compact star objects, or endpoints in the evolution of common astronomical formations, are great places to study the properties and composition of high-density matter. Recently, many compact stellar objects with high densities have been discovered \cite{lat1}, and often, these objects are confused for pulsars, rotating stars with strong magnetic fields, etc.

When a star reaches the end of its life, it uses up all of its nuclear energy sources, collapses owing to the force of its own gravity, and releases a massive amount of energy. One important astrophysical process that is in charge of the universe's structure construction is gravitational collapse. Black holes, neutron stars, white dwarfs, and other stellar remnants are the byproducts of stars collapsing due to gravity. The final product type of a gravitationally collapsing system is determined by the parent star system's initial mass configuration. A tiny or average-sized star may collapse to form a white dwarf, where the pressure from the electron degeneracy prevents the star from falling any further. But later, the collapse resumes and the white dwarf reaches the neutron star stage if it absorbs enough material from the surrounding environment to exceed the Chandrasekhar limit (greater than 1.39 solar mass). The neutron degeneracy pressure at this point prevents further collapse, but it can be overcome by re-accreting mass from the surrounding material. Subsequently, a simultaneous supernova event triggers another collapse, which culminates in a singularity. Thus, it is evident that the star's mass plays the most significant role during the entire collapsing sequence. A supermassive star's collapse, which begins with a massive starting mass (greater than 30 solar masses), seldom ends at any intermediate stage (white dwarf or neutron star) but eventually reaches a singularity. The gravitational collapse's astrophysical relevance sparks a lot of research on the subject. Oppenheimer and Snyder's groundbreaking research \cite{g1}, which examined the collapse of a dust cloud represented by a Schwarzschild shell and a Friedmann interior, set the stage for the entire endeavor in 1939. Tolman \cite{g2} and Bondi \cite{g3} investigated the collapse of an inhomogeneous dust distribution with a spherically symmetric distribution after that. Black holes emerged from each of these collapsing scenarios. Comprehensive analyses of gravitational collapse may be found in \cite{g4,g5}. The Cosmic Censorship Hypothesis (CCH), first put forth by Roger Penrose in 1969, \cite{g6} asserted that any singularity resulting from gravitational collapse would invariably be imprisoned inside an event horizon, named a black hole. This entailed totally hiding the singularity from outside observers so that no knowledge of its innards would seep into the outside world. The information loss paradox is the name given to this black hole phenomenon \cite{g7,g8,g9,g10}. However, there were many questions about the legitimacy of the CCH because there was not a complete confirmation of it. So the researchers started looking for evidence to refute the theory. To do that, it was now necessary to demonstrate that a collapsing process had provisions for the development of a naked singularity (NS) in addition to always producing a trapped surface \cite{g11,g12,g13,g14,g15,g16,g17}. No event horizon can occur for such a singularity, and information can flow continuously to and from the singularity. This immediately addresses the issue of information loss. Furthermore, our quantum gravity problem will gain momentum from the constant flow of data coming directly from the center of the singularity. Consequently, near the close of the 20th century, research on NS started with the publications of Eardly and Smarr \cite{g11} and Christodoulou \cite{g12}. Additional noteworthy NS works are extensively documented in the references \cite{g18,g19,g20}.

The spacetime outside of a spherically symmetric matter distribution with a constant mass was represented by the Schwarzschild metric. Since it is static, it is unable to simulate the spacetime of a realistic star. Vaidya provided a solution to this issue in his groundbreaking 1951 paper \cite{g21}, proposing a dynamic spacetime outside of a realistic star. The shining or radiating Schwarzschild metric is a fitting moniker for Vaidya's metric. Here, a time-dependent dynamic mass parameter took the place of the constant mass parameter in the Schwarzschild metric and served as the origin of the outgoing star radiations. The references \cite{g22,g23,g24,g25,g26,g27,g28,g29,g30,g31,g32} contain a variety of studies conducted in Vaidya spacetime.

It is estimated that dark matter, a sort of hypothetical material, accounts for 27$\%$ of the universe's mass-to-energy ratio. Though there are several possibilities from particle physics and supersymmetric string theory, like axions and wimps, there hasn't been a direct scientific finding of dark matter to date. Despite this, a bunch of recent experimental data points to the presence of dark matter in a number of astrophysical phenomena, including the formation of galactic rotation curves \cite{na1}, the dynamics of galaxy clusters \cite{na2}, and the cosmological scales of anisotropies found in the cosmic microwave background by PLANCK \cite{na3}. Dark matter is one of the greatest scientific mysteries of our time. Considering how little is known about its microphysical properties, numerous identities for it exist. Dark matter masses can range from $10^8$ solar mass $\simeq$ $10^{74}$ eV, which is the mass of DM in a compact galaxy, to $10^{-24}$ eV, which is determined by the largest possible Compton wavelength that can be contained within a dwarf galaxy. Over this range of masses, DM can be described as a particle, a wave/field, a macroscopic object, or a galaxy substructure, including topological defects and black holes. One practical strategy to deal with such great variation in potential is to take advantage of physical systems that have extraordinary diversity in their properties.

The quantum statistics of integer spin particles, or bosons, have been a key area of research in theoretical and experimental physics since Bose's conception \cite{na4} and Einstein's generalization \cite{na5,na6}. The bosonic systems' phase change to a condensed state is their most significant feature. This type of quantum bosonic system is known as a Bose-Einstein Condensate (BEC) because all of the particles are in the same quantum ground state. In both coordinate and momentum space, a system of this kind is physically typified by a strong peak over a wider distribution. The particles within a BEC exhibit mutual correlation and wavelength overlap. Many fundamental phenomena in condensed matter physics are thought to be understood in large part through the Bose-Einstein condensation process. One way to explain the superfluidity of low-temperature liquids, such as $^{3}$He, is to assume that a Bose-Einstein Condensation process occurs \cite{na7}. Bose-Einstein condensation has been observed and extensively researched in Earthly systems, and thus it is not immediately possible to rule out the idea that it also occurs in bosonic systems that exist at astrophysical or cosmological scales. It was therefore hypothesized that dark matter, which is thought to be a cold bosonic gravitationally constrained system and which is necessary to explain the dynamics of the neutral hydrogen clouds at great distances from the galaxy centers, could also exist in the form of a Bose-Einstein condensate \cite{na8,na9,na10,na11,na12,na13}. The features of Bose-Einstein Condensed Galactic Dark Matter Halos are systematically studied in \cite{sus1}. The presence of Bose-Einstein Condensed dark matter and its implications for astrophysics and cosmology have recently been extensively studied in the literature \cite{sus2,sus3,sus4,sus5,sus6,sus7,sus8,sus9,sus10,sus11,sus12}. Furthermore, the notion that some types of BEC could arise in neutron stars has long been entertained \cite{sus13}. Inspired by the aforementioned, we are motivated to study the gravitational collapse of a stellar system composed of Bose-Einstein condensate dark matter having a linear equation of state.

The underlying gravity theory will be considered as GR. Unless otherwise specified, we will use natural units
($c=8\pi G=1$) throughout this paper, and the Latin indices will run from 0 to 3. We will use the signature ($-,+,+,+$). The Riemann tensor is defined by
\begin{equation}
R^{a}_{bcd}=\Gamma^{a}_{bd,c}-\Gamma^{a}_{bc,d}+\Gamma^{e}_{bd}\Gamma^{a}_{ce}-\Gamma^{e}_{bc}\Gamma^{a}_{de}
\end{equation}
where $\Gamma^{a}_{bd}$ are the Christoffel symbols defined by,
\begin{equation}
\Gamma^{a}_{bd}=\frac{1}{2}g^{ae}\left(g_{be,d}+g_{ed,b}-g_{bd,e}\right)
\end{equation}
where $g_{ab}$ is the metric tensor. The Ricci tensor is obtained by contracting the first and the third indices of the Riemann tensor
\begin{equation}
R_{ab}=g^{cd}R_{cadb}    
\end{equation}

The Einstein-Hilbert action when matter is present is given by,
\begin{equation}
S=\frac{1}{2}\int d^{4}x\sqrt{-g}\left(R-2\Lambda-2\mathcal{L}_{m}\right)
\end{equation}
where $R$ is the Ricci scalar, $\Lambda$ is the cosmological constant, and $\mathcal{L}_{m}$ is the matter Lagrangian. Taking variation of the above equation we get the Einstein field equations,
\begin{equation}\label{field}
G_{ab}+\Lambda g_{ab}=T_{ab}
\end{equation}
where $G_{ab}$ is the Einstein tensor, and $T_{ab}$ is the energy-momentum tensor (EMT). The paper is organized as follows: The generalized Vaidya spacetime in the presence of BEC will be studied in section 2. In section 3 the gravitational collapse mechanism will be explored and the nature of singularities will be studied. Section 4 is dedicated to the study of the junction conditions. The paper will end with a discussion and conclusion in section 5.

\section{Generalized Vaidya spacetime}
The most general spherically symmetric line element for any combination of Type-I or Type-II matter fields is given by,
\begin{equation}
ds^{2}=-e^{2\psi(\nu,r)}\left[1-\frac{2m(\nu,r)}{r}\right]d\nu^{2}+2\epsilon~ e^{\psi(\nu,r)}d\nu dr+r^{2}\left(d\theta^{2}+sin^{2}\theta d\phi^{2}\right), ~~~~~~~(\epsilon=\pm 1)
\end{equation}
where $\nu$ is the null coordinate and $r$ is the radial coordinate. $m(\nu,r)$ is the mass function corresponding to the gravitational energy within a given radius $r$. For $\epsilon=+1$ the Eddington advanced time is represented by the null coordinate $\nu$. Here $r$ decreases towards the future along a ray $\nu=$constant, which represents ingoing null congruence. On the other hand, $\epsilon=-1$ represents outgoing null congruence. The particular blend of matter fields that makes $\psi(\nu,r)=0$ gives the generalized Vaidya geometry. Moreover, since we are considering a collapsing scenario, we take $\epsilon=+1$. So we get the line element of the form \cite{g21},
\begin{equation}\label{vd0}
ds^{2}=-\left[1-\frac{2m(\nu,r)}{r}\right]d\nu^{2}+2 d\nu dr+r^{2}\left(d\theta^{2}+sin^{2}\theta d\phi^{2}\right)
\end{equation}
The non-zero components of the Ricci tensor can be written as,
\begin{eqnarray}
R_{\nu\nu}=\frac{-\left(r-2m(\nu,r)\right)m''(\nu,r)+2\dot{m}(\nu,r)}{r^2}, ~~~~~~~~ R_{\nu r}=R_{r\nu}=\frac{m''(\nu,r)}{r},\\ \nonumber R_{\theta\theta}=2m'(\nu,r),~~~~~~~ 
R_{\phi\phi}=2m'(\nu,r) \sin^{2}{\theta}
\end{eqnarray}
where a 'dash' represents derivative with respect to $r$ and a 'dot' represents derivative with respect to $\nu$.
The Ricci scalar is given by,
\begin{equation}
R=\frac{2m''(\nu,r)}{r}+\frac{4m'(\nu,r)}{r^{2}}
\end{equation}
The non-zero components of the Einstein tensor are given by,
\begin{eqnarray}
G_{\nu\nu}=-\frac{2\left[\left(r-2m(\nu,r)\right)m'(\nu,r)+r\dot{m}(\nu,r)\right]}{r^{3}}, ~~~~~~~ G_{\nu r}=G_{r\nu}=-\frac{2m'(\nu,r)}{r^{2}},\\ \nonumber
G_{\theta\theta}=-rm''(\nu,r),~~~~~~~ G_{\phi\phi}=-rm''(\nu,r)\sin^{2}{\theta}
\end{eqnarray}
The corresponding energy-momentum tensor can be given in the form,
\begin{equation}
T_{\mu\nu}=T_{\mu\nu}^{(n)}+T_{\mu\nu}^{(m)}
\end{equation}
where $T_{\mu\nu}^{(n)}$ and $T_{\mu\nu}^{(m)}$ are the contributions from Vaidya null radiation and matter respectively. Here these components of the EMT are given by,
\begin{equation}
T_{\mu\nu}^{(n)}=\sigma l_{\mu}l_{\nu}, ~~~ T_{\mu\nu}^{(m)}=\left(\rho+p\right)\left(l_{\mu}\eta_{\nu}+l_{\nu}\eta_{\mu}\right)+pg_{\mu\nu}
\end{equation}
In the above definitions $l_{\mu}$ and $\eta_{\mu}$ are two null vectors given by,
\begin{equation}
l_{\mu}=\delta_{\mu}^{0}, ~~~ \eta_{\mu}=\frac{1}{2}\left[1-\frac{2m(\nu,r)}{r}\right]\delta_{\mu}^{0}-\delta_{\mu}^{1}
\end{equation}
where $l_{\mu}l^{\mu}=\eta_{\mu}\eta^{\mu}=0$ and $l_{\mu}\eta^{\mu}=-1$.

$T_{\mu\nu}^{(n)}$ can be considered to be the matter field that moves along the null hypersurfaces $\nu=$constant, while $T_{\mu\nu}^{(m)}$ represents matter moving along time-like curves. If we consider $\rho=p=0$, then we deduce the Vaidya solution with $m=m(\nu)$.

The non-vanishing components of the energy-momentum tensor will be as follows
\begin{eqnarray*}
T_{\nu\nu}=\sigma+\rho\left(1-\frac{2m(\nu,r)}{r}\right),&&
~~T_{\nu r}=-\rho \\
\end{eqnarray*}
\begin{eqnarray}\label{energymomentum}
T_{\theta\theta}=pr^2, && ~~T_{\phi\phi}=pr^2 \sin^2\theta
\end{eqnarray}

\subsection{Dark matter in the form of Bose-Einstein condensate}
The literature demonstrates that because compact stars have high densities, scientists are always attempting to determine the precise makeup of the matter distribution. Recently, a number of researchers presented different models for compact stars with DM composition in order to achieve this goal \cite{bec1, bec2, bec3, bec4, bec5}. From these perspectives, we offer a viable and relatively new model for compact stars generated by DM by taking into account the Bose-Einstein DM density profile and a linear equation of state (EoS) \cite{bec6}.
The density profile of Bose-Einstein condensate (BEC) is given by \cite{bec6}
\begin{equation}\label{den}
\rho=\rho(r)=\frac{a}{kr}\sin{(kr)}
\end{equation}
where $a(km^{-1})$ and $k$ are non-zero positive constants. The linear EoS is given by,
\begin{equation}\label{eoslin}
p=p(r)=\alpha \rho(r)-\beta
\end{equation}

According to research by Chavanis and Harko \cite{bec7}, a sizable portion of the matter in some astronomical compact objects might be in the form of Bose-Einstein condensate. The existence of Bose-Einstein condensates in neutron stars in helium white dwarf stars was covered by Pethik et al. \cite{bec8} and Mosquera et al. \cite{bec9}, respectively. Moreover, Li et al. \cite{bec10} investigated the matter distributions that may be produced from the Bose-Einstein condensation of dark matter. We introduced this model as a result of all this important research and the non-singular feature of the density profile mentioned above. Using Eqn.(\ref{den}) in Eqn.(\ref{eoslin}) we get
\begin{equation}
p=\frac{a\alpha}{kr}\sin{(kr)}-\beta
\end{equation}

\subsection{Field Equations}
Now we will proceed to form the field equations of generalized Vaidya spacetime filled with Bose-Einstein DM. We will start by using the field equations given in Eqn.(\ref{field}) and put all the tensor components that we have calculated previously. Below we report the different components of the field equations.\\

The $(\nu\nu)$ component is given by,
\begin{equation}
2\left[\left(r-2m\right)\frac{\partial m}{\partial r}+r\frac{\partial m}{\partial \nu}\right]+\left(\Lambda+\rho\right)\left(-r+2m\right)r^{2}-\sigma r^{3}=0
\end{equation}

The $(\nu r)$ component is given by,
\begin{equation}
2\frac{\partial m}{\partial r}-\left(\Lambda+\rho\right)r^{2}=0
\end{equation}

The $(\theta\theta)$ component is given by,
\begin{equation}
r\frac{\partial^{2} m}{\partial r^{2}}-\left(\Lambda-p\right)r^{2}=0
\end{equation}

The $(\phi\phi)$ component is given by,
\begin{equation}
r\frac{\partial^{2} m}{\partial r^{2}}-\left(\Lambda-p\right)r^{2}=0, ~~~~~~\theta\neq m\pi, ~~m\in \mathbb{Z}
\end{equation}
It should be noted that the $(\theta\theta)$ and $(\phi\phi)$ components produce similar field equations with the exception of the condition on $\theta$ for the latter.

Now we proceed to introduce the Bose-Einstein DM density into the field equations. The $(\nu\nu)$ component becomes,
\begin{equation}\label{f1}
2\left[\left(r-2m\right)\frac{\partial m}{\partial r}+r\frac{\partial m}{\partial \nu}\right]+\left[\Lambda+\frac{a}{kr}\sin{(kr)}\right]\left(-r+2m\right)r^{2}-\sigma r^{3}=0
\end{equation}
The $(\nu r)$ component takes the shape,
\begin{equation}\label{f2}
2\frac{\partial m}{\partial r}-\left[\Lambda+\frac{a}{kr}\sin{(kr)}\right]r^{2}=0
\end{equation}
The $(\theta\theta)$ and $(\phi\phi)$ components are given by,
\begin{equation}\label{f3}
r\frac{\partial^{2} m}{\partial r^{2}}-\left[\Lambda-\frac{a\alpha}{kr}\sin{(kr)}+\beta\right]r^{2}=0
\end{equation}
It is quite clear that the above field equations are differential equations in terms of the Vaidya mass function $m(\nu,r)$. In the next section, we will investigate the solutions of the above differential equations.

\subsection{Solution of the system}
In this section, we will find the solution to the differential equations set up in the previous section. Looking at the equations it seems that Eqns.(\ref{f2}) and (\ref{f3}) are relatively simpler than Eqn.(\ref{f1}). So we first proceed to seek solutions of Eqns.(\ref{f2}) and (\ref{f3}). Solving Eqn.(\ref{f2}) we get,
\begin{equation}\label{sol1}
m(\nu,r)=\frac{a\sin{kr}}{2k^3}-\frac{ar\cos{kr}}{2k^2}+\frac{\Lambda r^{3}}{6}+f_{1}(\nu)
\end{equation}
where $f_{1}(\nu)$ is an arbitrary function of $\nu$.

Solving Eqn.(\ref{f3}) we get,
\begin{equation}\label{sol2}
m(\nu,r)=\frac{a\alpha\sin{kr}}{k^3}+\frac{\beta r^3}{6}+\frac{\Lambda r^{3}}{6}+rf_{2}(\nu)+f_{3}(\nu)
\end{equation}
where $f_{2}(\nu)$ and $f_{3}(\nu)$ are arbitrary functions of $\nu$.

Unfortunately, we do not have a general solution for the Eqn.(\ref{f1}). But in the limit $r\rightarrow 0$, i.e., at the center of the stellar system we have the following solution,
\begin{equation}\label{sol3}
m(\nu,r)=\pm\left[\frac{\left(a+\Lambda\right)r+4f_{4}(\nu)}{2}\right]^{1/2}
\end{equation}
where $f_{4}(\nu)$ is an arbitrary function of $\nu$. This solution is a trivial one since it does not preserve any contributions of the Bose-Einstein DM density profile. The solution contains two branches, the first one is positive and the other one is negative. Although negative mass is not possible for normal matter, since we are dealing with a highly exotic form of matter, there might always be a possibility. 

In the subsequent analysis we will be working with the solutions given in Eqns.(\ref{sol1}) and (\ref{sol2}) because they preserve the contributions coming from the Bose-Einstein DM density profile, which is the main feature under study in this work. So we would like to discuss these two solutions in detail before we go on to use them in the latter sections. First, let us discuss the two solutions comparatively. The solution given in Eqn.(\ref{sol1}) contains four terms whereas the solution given in Eqn.(\ref{sol2}) contains five terms. For $\alpha=1/2$ the first terms in both the solutions coincide. Now if we expand the $\cos{kr}$ in the second term of Eqn.(\ref{sol1}) in series we get $-\frac{ar}{2k^2}\left(1-\frac{k^{2}r^{2}}{2}+.....\right)$. Taking only the first two terms of the infinite series we get $\frac{a}{4}r^{3}-\frac{ar}{2k^{2}}$. These two terms coincide with the second and the fourth terms of the solution in Eqn.(\ref{sol2}) for $\beta=3a/2$ and $f_{2}(\nu)=-a/2k^{2}$ respectively. The third and the last terms in both solutions coincide. So it is clear that the solutions given in Eqns.(\ref{sol1}) and (\ref{sol2}) are almost identical (at least approximately) and we can actually work with either of the two. It is better to work with the solution in Eqn.(\ref{sol2}) because it has more terms and more free parameters, bringing in the complete flavor of the Bose-Einstein condensate in the generalized Vaidya spacetime. Moreover, this solution has two arbitrary functions in terms of $\nu$ which increases the degree of freedom. Now as $r\rightarrow 0$, we have $m(\nu,r)\rightarrow f_{3}(\nu)$, which is the self-similar Vaidya spacetime. So at the center of the stellar system, the geometry is reduced to a self-similar Vaidya geometry. In the limit $r\rightarrow \infty$ we have $m(\nu,r)\rightarrow \infty$, which physically does not make any sense. But it obviously it shows that the mass of the star increases as we move away from the center. At the Schwarzschild radius $r_{s}$ we have $m(\nu,r)|_{r=r_{s}}=\frac{a\alpha\sin{kr_{s}}}{k^3}+\frac{\beta r_{s}^3}{6}+\frac{\Lambda r_{s}^{3}}{6}+r_{s}f_{2}(\nu)+f_{3}(\nu)$. This should be the mass of the Schwarzschild black hole formed (if any). So the generalized Vaidya spacetime with Bose-Einstein DM density is given by,
\begin{equation}\label{vd1}
ds^{2}=-\left[1-\frac{2a\alpha\sin{kr}}{rk^3}-\frac{\left(\beta+\Lambda\right) r^2}{3}-2h(\nu)-\frac{2g(\nu)}{r}\right]d\nu^{2}+2 d\nu dr+r^{2}\left(d\theta^{2}+sin^{2}\theta d\phi^{2}\right)
\end{equation}
where $f_{2}(\nu)$ and $f_{3}(\nu)$ are renamed as $h(\nu)$ and $g(\nu)$ respectively to simplify the nomenclature. We generalize the above solution further to introduce the contribution from the null radiation $\sigma$. The generalized form becomes,
\begin{equation}\label{vd2}
ds^{2}=-\left[1-\frac{2a\alpha\sin{kr}}{rk^3}-\frac{\left(\beta+\Lambda\right) r^2}{3}-2h(\nu)-\frac{2g(\nu)}{r}-\frac{2\sigma}{r}\right]d\nu^{2}+2 d\nu dr+r^{2}\left(d\theta^{2}+sin^{2}\theta d\phi^{2}\right)
\end{equation}
Here we see that in Eqn.(\ref{vd1}) we have the arbitrary function $h(\nu)$ which may be written as $h(\nu)=f_{4}(\nu)+constant$. The constant $\sigma$ considered above can easily be adjusted with this constant in the function $h(\nu)$. So mathematically there is no problem in introducing $\sigma$, which gives the contribution coming from the Vaidya null radiation. In the next section, we will study a collapsing scenario in this spacetime and probe the end state of the collapse.

\section{Gravitational collapse: nature of singularity}
We know that $ds^{2}=0$ provides the directions that light rays move for every relativistic metric. The investigation of light cones in the context of the special theory of relativity makes this clear. Here, we would investigate the possibility that the central singularity's core emits light rays (radial null geodesics). The equation for such geodesics can be found from the metric in Eqn.(\ref{vd0}) by equating $ds^{2}=0$ and $d\Omega_{2}^{2}=d\theta^{2}+sin^{2}\theta d\phi^{2}=0$. Performing these operations we get,
\begin{equation}
\frac{d\nu}{dr}=\frac{2r}{r-2m(\nu,r)}
\end{equation}
The above differential equation has a singularity at $r=0$, $\nu=0$. Since all mathematical and physical structures completely collapse at the singularity, we must investigate the limiting behavior of the trajectories as they get closer to the singularity. For this study, we consider a parameter $\chi=\nu/r$. As one travels along the trajectory of the radial null geodesic towards the singularity at $r=0, ~\nu=0$, we will examine the limiting behavior of the parameter $\chi$. Suppose the limiting value of $\chi$ at $r=0, ~\nu=0$ is given by $\chi_{0}$, then by L'Hospital's rule we have
\begin{eqnarray}\label{X0}
\begin{array}{c}
\chi_{0}\\\\
{}
\end{array}
\begin{array}{c}
=\lim \chi \\
\begin{tiny}\nu\rightarrow 0\end{tiny}\\
\begin{tiny}r\rightarrow 0\end{tiny}
\end{array}
\begin{array}{c}
=\lim \frac{\nu}{r} \\
\begin{tiny}\nu\rightarrow 0\end{tiny}\\
\begin{tiny}r\rightarrow 0\end{tiny}
\end{array}
\begin{array}{c}
=\lim \frac{d\nu}{dr} \\
\begin{tiny}\nu\rightarrow 0\end{tiny}\\
\begin{tiny}r\rightarrow 0\end{tiny}
\end{array}
\begin{array}{c}
=\lim \frac{2r}{r-2m(\nu,r)} \\
\begin{tiny}\nu\rightarrow 0\end{tiny}~~~~~~~~~~~~\\
\begin{tiny}r\rightarrow 0\end{tiny}~~~~~~~~~~~~
 {}
\end{array}
\end{eqnarray}
An algebraic equation in terms of $\chi_0$ can be constructed from the above limit by using the mass term that we previously obtained. We are going to search for the roots of this equation, which are actually the directions of the tangents to the outward geodesics. As they reflect the actual information exchange with the outside observers via escaping radiation, the true roots of the equation should, in theory, be the only ones that concern us. Any positive real root of the equation in our mathematical setup will show the gradient of the tangent to an outgoing null geodesic. Thus, the formation of a naked singularity (NS) as a result of the stellar collapse is indicated by the presence of real positive roots in the algebraic equation that will be produced.

A single wavefront emanating from the core singularity and illuminating the external observer is correlated with a single null geodesic breaking away from the singularity. In that scenario, the singularity would become visible to a distant observer instantaneously, prior to any obstruction to their field of view. There will be a locally naked singularity as a result. In this case, the singularity momentarily becomes visible, physically delaying the creation of the trapped surface for a short while. However, a complete exchange of data and knowledge between the distant spectator and the singularity may require more than this brief exposure. Therefore, the singularity must be visible for a longer period of time in order to provide a thorough trade-off of information. This can happen if the central singularity emits a bundle of null geodesics, rendering it globally naked. With our theoretical framework, it is quite easy to investigate this by counting the number of real positive roots that may be obtained from the algebraic equation. The aforementioned mathematical formulation was initially put forth by Joshi, Singh, and Dwivedi in a number of their publications \cite{joshi1, joshi2, joshi3, joshi4} to investigate the nature of singularity created as the end state of a stellar collapse. 

The threshold mass at which black hole creation would be avoided depends on a number of factors, including the physical properties of the collapsing object, the equation of state of the matter involved, and the specifics of the collapse process. It's important to keep in mind that it is challenging to produce precise quantitative estimates due to the complexity of the problem and the variety of scenarios that could lead to the birth of a black hole. Still, a general summary of certain key concepts can be given. The Chandrasekhar limit \cite{chandra1} is the critical mass beyond which electron degeneracy pressure is unable to counterbalance gravity for a non-rotating, non-charged star maintained by electron degeneracy pressure, resulting in gravitational collapse. Approximately 1.44 $M_{\odot}$, where $M_{\odot}$ is one solar mass, is this limit. The collapse of more massive objects, such as neutron stars, is resisted by the pressure of neutron degeneracy. The maximum mass that a neutron star can have before collapsing into a black hole is known as the Tolman-Oppenheimer-Volkof (TOV) limit \cite{tolman1, oppo1}. This limit is around 2.16 $M_{\odot}$, however, the precise value is unknown because of problems with the equation of state for stuff such as neutrons. The case of big stars involves a unique mechanism. For stars with masses between 100 and 250 $M_{\odot}$, pair formation can cause a sharp reduction in pressure, which can lead to a pair-instability supernova. In certain cases, the star is completely disrupted, which prevents the formation of black holes. In certain cases, black holes can form directly ahead of schedule, negating the need for a supermassive star to go through an intermediate stage. This could happen in densely populated, low-metallicity regions where nuclear burning and radiation pressure don't prevent the collapse. The mass threshold for stars that collapse straight into black holes may be in the range of 102 -103 $M_{\odot}$, though this is not a certain value. It is important to emphasize that these are simply estimates and that the actual mass thresholds will differ based on the specific conditions and features of the collapsing object. Furthermore, astrophysical observations enable us to continuously refine our understanding of these processes. The aforementioned numerical approximations are given with standard general relativity as the background theory of gravity.  In the case of our study, we have considered Bose-Einstein DM as the constituent of the stellar system with the background gravity as GR. This is expected to alter the collapsing scenario considerably and hence the end state of the collapse. The exotic nature of condensate will have a profound effect on the collapsing mechanism and we will explore how does it alter the end state of the collapse.

Using the solution in Eqn.(\ref{vd2}) in Eqn.(\ref{X0}) we have,
\begin{eqnarray}\label{X01}
\chi_{0}=
\begin{array}llim\\
\begin{tiny}\nu\rightarrow 0\end{tiny}\\
\begin{tiny}r\rightarrow 0\end{tiny}
\end{array}
\left[\frac{2r}{r-\frac{2a\alpha\sin{kr}}{k^3}-\frac{\left(\beta+\Lambda\right) r^3}{3}-2rh(\nu)-2g(\nu)-2\sigma}\right]
\end{eqnarray}
Now to obtain self-similar solutions we consider $h(\nu)=h_{0}\nu$ and $g(\nu)=g_{0}\nu$. Similarly, we consider the null-radiation parameter as $\sigma(\nu)=\sigma_{0}\nu$. Here $h_{0}$, $g_{0}$, and $\sigma_{0}$ are constants depending on the initial conditions of the collapse. So using the above considerations we get from Eqn.(\ref{X01})
\begin{eqnarray}\label{X02}
\frac{\chi_{0}}{2}=
\begin{array}llim\\
\begin{tiny}\nu\rightarrow 0\end{tiny}\\
\begin{tiny}r\rightarrow 0\end{tiny}
\end{array}
\left[\frac{1}{1-\frac{2a\alpha}{k^2}\left(\frac{\sin{kr}}{kr}\right)-\frac{\left(\beta+\Lambda\right) r^2}{3}-2h_{0}\nu-2g_{0}\left(\frac{\nu}{r}\right)-2\sigma_{0}\left(\frac{\nu}{r}\right)}\right]
\end{eqnarray}
Calculating the above limit and using the definition given in Eqn.(\ref{X0}) we get
\begin{equation}
2\left(g_{0}+\sigma_{0}\right)\chi_{0}^{2}+\left(\frac{2a\alpha}{k^2}-1\right)\chi_{0}+2=0
\end{equation}
This is a quadratic equation in terms of the collapse parameter $\chi_{0}$ and has real roots for $\Delta=\left(1-\frac{2a\alpha}{k^2}\right)^2-16\left(g_{0}+\sigma_{0}\right)\geq 0$. Solving the above equation we get
\begin{equation}\label{roots}
\chi_{0_{1,2}}=\frac{1}{4\left(g_{0}+\sigma_{0}\right)}\left[1-\frac{2a\alpha}{k^2}\pm \sqrt{\left(1-\frac{2a\alpha}{k^2}\right)^2-16\left(g_{0}+\sigma_{0}\right)}\right]   
\end{equation}
We consider the positive root as $\chi_{0_{1}}$ and the negative root as $\chi_{0_{2}}$. The nature and signature of the above roots will depend on the parameter space that we use which will further depend on the initial conditions of the collapse. Table (\ref{tab1}) shows the different types of end states of collapse possible with their corresponding conditions.

\begin{table}
\caption{The table shows the different outcomes of the collapse for different signatures of the roots}
    \centering
    \begin{tabular}{||c|c|c|c||}
    \hline
        Nature of roots~~ & End state of collapse &Type  \\[1ex]
        \hline
$\chi_{0_{1}}>0$, ~$\chi_{0_{2}}<0$~~ & NS & Local  \\[1ex]
\hline
$\chi_{0_{1}}<0$, ~$\chi_{0_{2}}>0$~~ & NS  & Local  \\[1ex]
\hline
$\chi_{0_{1}}>0$, ~$\chi_{0_{2}}>0$~~ & NS  & Global  \\[1ex]
\hline
$\chi_{0_{1}}<0$, ~$\chi_{0_{2}}<0$~~ & BH  & --  \\[1ex]
\hline
Imaginary roots ($\Delta<0$)~~ & BH  & --  \\[1ex]
\hline
    \end{tabular}
    
    \label{tab1}
\end{table}

Now in order to understand the effect of the parameter space on the nature of roots we generate plots of the roots $\chi_{0_{1}}$ and $\chi_{0_{2}}$ for different parametric perturbations.

\begin{figure}[hbt!]
\begin{center}
\includegraphics[height=2.5in]{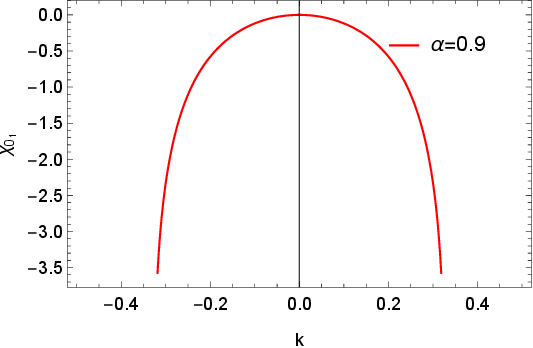}
\caption{Plot of the collapse parameter $\chi_{0_{1}}$ against the BEC parameter $k$ for EoS parameter $\alpha=0.9$. The other parameters are considered as $g_{0}=0.01$,~ $\sigma_{0}=0.02$,~$a=0.1$.}
\label{figep1}
\end{center}
\end{figure}

\begin{figure}[hbt!]
\begin{center}
\includegraphics[height=2.5in]{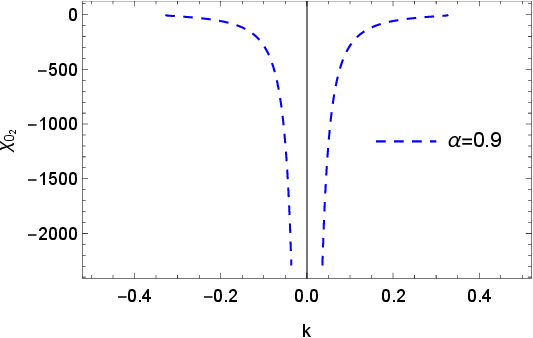}
\caption{Plot of the collapse parameter $\chi_{0_{2}}$ against the BEC parameter $k$ for EoS parameter $\alpha=0.9$. The other parameters are considered as $g_{0}=0.01$,~ $\sigma_{0}=0.02$,~$a=0.1$.}
\label{figep2}
\end{center}
\end{figure}

In Fig.(\ref{figep1}) we have generated the plot of the collapse parameter $\chi_{0_{1}}$ against the BEC parameter $k$ for EoS parameter $\alpha=0.9$. We see that the entire trajectory remains in the negative levels. Similarly in Fig.(\ref{figep2}) the collapse parameter $\chi_{0_{1}}$ is plotted against the BEC parameter $k$ for EoS parameter $\alpha=0.9$. Here also we see that the entire trajectory resides in the negative level. So for the EoS parameter $\alpha=0.9$ we see that both the roots of the Eqn.(\ref{roots}) are negative indicating the formation of BH as the end state of collapse. In fact for all $\alpha>0$ we get BH as the result of collapse. We know that $\alpha$ is the EoS parameter, whose value determines the nature of matter filling the star. Here $\alpha$ plays the role of a barotropic parameter. So $\alpha>0$ signifies ordinary matter, which we see from our analysis results in a BH. Moreover, it should be noted that the BEC parameter does have a significant effect on the collapse. We see that around $k=0$, there is a turnaround in the trajectories of the roots, showing the existence of a maxima/minima. For the first root, there is a maximum at $k=0$, and for the second root, there is a minimum. Although it seems that the second root diverges around $k=0$, that is not the case. It is simply an effect of the choice of the parameter space.

\begin{figure}[hbt!]
\begin{center}
\includegraphics[height=2.5in]{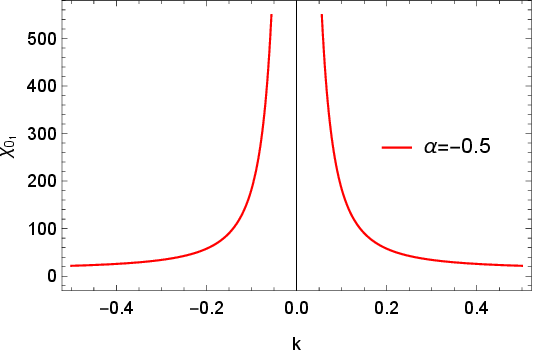}
\caption{Plot of the collapse parameter $\chi_{0_{1}}$ against the BEC parameter $k$ for EoS parameter $\alpha=-0.5$. The other parameters are considered as $g_{0}=0.01$,~ $\sigma_{0}=0.02$,~$a=0.1$.}
\label{figep3}
\end{center}
\end{figure}

\begin{figure}[hbt!]
\begin{center}
\includegraphics[height=2.5in]{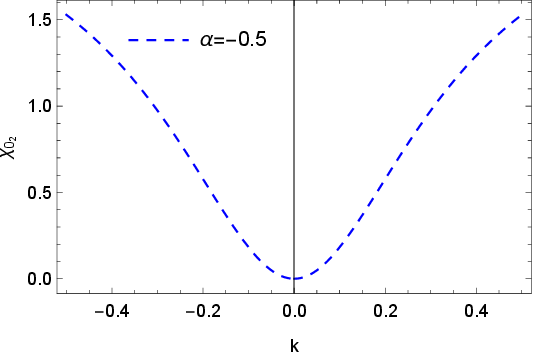}
\caption{Plot of the collapse parameter $\chi_{0_{2}}$ against the BEC parameter $k$ for EoS parameter $\alpha=-0.5$. The other parameters are considered as $g_{0}=0.01$,~ $\sigma_{0}=0.02$,~$a=0.1$.}
\label{figep4}
\end{center}
\end{figure}

In Figs.(\ref{figep3}) and (\ref{figep4}) we have generated the plots of the collapse parameter $\chi_{0_{1}}$ and $\chi_{0_{2}}$ respectively against the BEC parameter $k$ for EoS parameter $\alpha=-0.5$. We see that both the roots are positive in this case. In fact, for all $\alpha<0$, we get positive values of both the roots. Hence here the collapse results in a global NS. Generally, $\alpha<-1/3$ signifies exotic matter. So we see that if the star is made up of an exotic form of BEC then the resulting singularity from the gravitational collapse is globally naked, which violates the cosmic censorship hypothesis. Just like the first two figures, here also we have maximum and minimum around the $k=0$ mark. Here the first root has a maximum and the second root has a minimum at $k=0$. A transition in the direction of the trajectories may correspond to a phase transition inside the dark matter, which is really important information regarding the model.

\begin{figure}[hbt!]
\begin{center}
\includegraphics[height=3in]{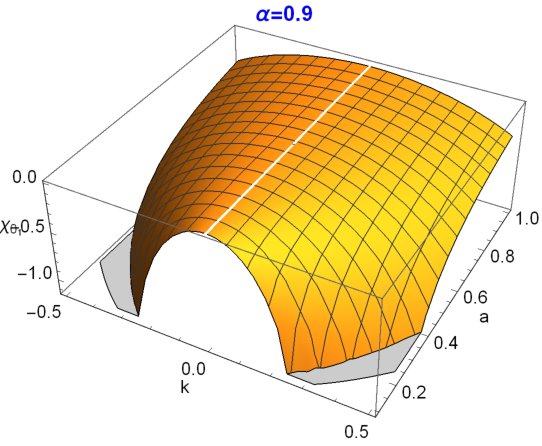}
\caption{3D-Plot of the collapse parameter $\chi_{0_{1}}$ aginst the $k-a$ plane for EoS parameter $\alpha=0.9$. The other parameters are considered as $g_{0}=0.01$,~ $\sigma_{0}=0.02$.}
\label{figep5}
\end{center}
\end{figure}

\begin{figure}[hbt!]
\begin{center}
\includegraphics[height=3in]{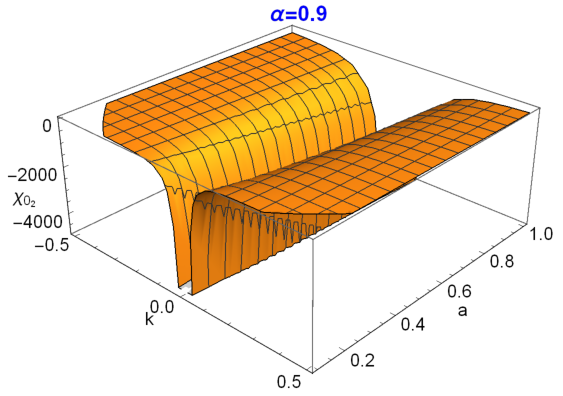}
\caption{3D-Plot of the collapse parameter $\chi_{0_{2}}$ aginst the $k-a$ plane for EoS parameter $\alpha=0.9$. The other parameters are considered as $g_{0}=0.01$,~ $\sigma_{0}=0.02$.}
\label{figep6}
\end{center}
\end{figure}

\begin{figure}[hbt!]
\begin{center}
\includegraphics[height=3in]{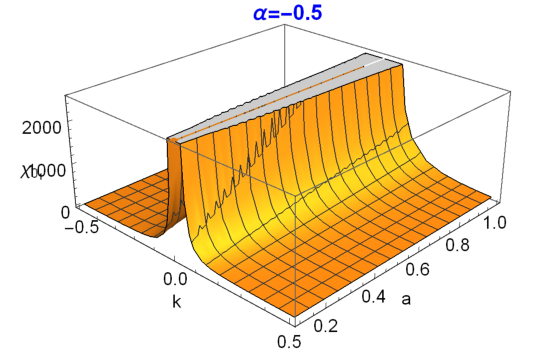}
\caption{3D-Plot of the collapse parameter $\chi_{0_{1}}$ aginst the $k-a$ plane for EoS parameter $\alpha=-0.5$. The other parameters are considered as $g_{0}=0.01$,~ $\sigma_{0}=0.02$.}
\label{figep7}
\end{center}
\end{figure}

\begin{figure}[hbt!]
\begin{center}
\includegraphics[height=3in]{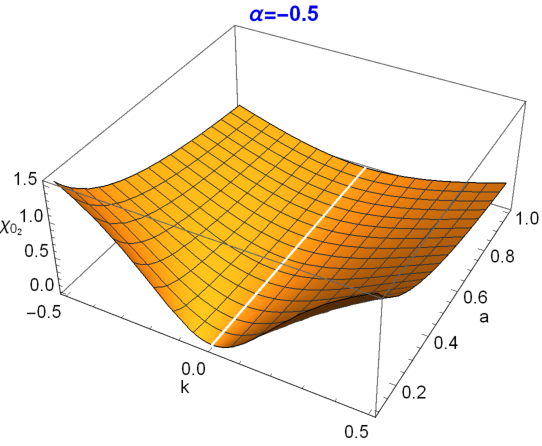}
\caption{3D-Plot of the collapse parameter $\chi_{0_{2}}$ aginst the $k-a$ plane for EoS parameter $\alpha=-0.5$. The other parameters are considered as $g_{0}=0.01$,~ $\sigma_{0}=0.02$.}
\label{figep8}
\end{center}
\end{figure}

To get a more accurate idea of the effect of the parameter space on the collapse parameter, we plot it simultaneously against both the BEC parameters $k$ and $a$ in 3-dimensional plots. In Figs.(\ref{figep5}) and (\ref{figep6}) we plot $\chi_{0_{1}}$ and $\chi_{0_{2}}$ respectively for $\alpha=0.9$. It is seen that both the roots remain in the negative levels confirming our find in Figs.(\ref{figep1}) and (\ref{figep2}). The resulting singularity, as discussed earlier, is shrouded by an event horizon and is a BH. Similarly in Figs.(\ref{figep7}) and (\ref{figep8}) we have generated similar plots of the collapse parameter for $\alpha=-0.5$. It is found that the roots remain in positive levels confirming our find in Figs.(\ref{figep3}) and (\ref{figep4}). So in this case we get global NS as the end state of the collapse which violates the cosmic censorship hypothesis. In these 3D plots, the transitions around $k=0$ are more prominently visible. These transitions are expected to be the consequence of phase transitions inside the BEC dark matter model. This can be due to a change in the chemical makeup or a structural change. It can also be due to a change in the different kinds of magnetic ordering in the matter.

\subsection{Strength of the singularity: curvature growth near the singularity}
The gravitational strength of a singularity is determined by the curvature imposed on the surrounding spacetime. It is an estimation of the singularity's destructive potential for any material that it comes into contact with. The existence of singularities in the theoretical framework of most theories of gravity has rendered them completely unworkable. There have been a number of theoretical approaches to the eradication of these singularities put out in the literature over the years, but they are essentially impractical and in no way acceptable. It is well known that singularities are voids in the otherwise smooth and continuous spacetime. When there is a weak singularity, the crater is shallow and spacetime can continue through the singularity without breaking. This concept has mathematical analogies to a removable discontinuity, which can restore spacetime's continuity at a singularity. We can see from the arguments above that there is sufficient reason to find out if a singularity is strong or weak. A curvature singularity will be strong, according to Tipler \cite{tip1}, if any item approaching it or colliding with it is compressed to zero volume. Ref.\cite{tip1} provides the following criteria for a strong singularity:
\begin{eqnarray}\label{tipu}
\begin{array}{c}
S=\lim \tau^{2}\psi \\
\begin{tiny}\tau\rightarrow 0\end{tiny}\\
\end{array}
\begin{array}{c}
=\lim \tau^{2}R_{ab}K^{a}K^{b}>0 \\
\begin{tiny}\tau\rightarrow 0\end{tiny}\\
\end{array}
\end{eqnarray}
where $R_{ab}$ is the Ricci tensor, $\psi$ is a scalar given by the relation $\psi=R_{ab}K^{a}K^{b}$, where
$K^{a}=dx^{a}/d\tau$ represents the tangent to the non
spacelike geodesics at the singularity and $\tau$ is the affine parameter. Mkenyeleye et al. \cite{mken1} have shown that,
\begin{eqnarray}\label{maha}
\begin{array}{c}
S=\lim \tau^{2}\psi \\
\begin{tiny}\tau\rightarrow 0\end{tiny}\\
\end{array}
\begin{array}{c}\label{stren}
=\frac{1}{4}\chi_{0}^{2}\left(2\dot{m_{0}}\right) \\
\begin{tiny}~\end{tiny}\\
\end{array}
\end{eqnarray}
where
\begin{eqnarray}
\begin{array}{c}
m_{0}=\lim~ m(\nu,r) \\
\begin{tiny}\nu\rightarrow 0\end{tiny}\\
\begin{tiny}r\rightarrow 0\end{tiny}
\end{array}
\end{eqnarray}
and
\begin{eqnarray}\label{massd}
\begin{array}{c}
\dot{m_{0}}=\lim \frac{\partial}{\partial~\nu}\left(m(\nu,r)\right) \\
\begin{tiny}\nu\rightarrow 0\end{tiny}\\
\begin{tiny}r\rightarrow 0\end{tiny}
\end{array}
\end{eqnarray}

It has also been demonstrated in ref.\cite{mken1} that the relationship between $\chi_{0}$ and the mass limiting values is provided by
\begin{equation}\label{xmass}
\chi_{0}=\frac{2}{1-2m_{0}'-2\dot{m_{0}}\chi_{0}}
\end{equation}
where
\begin{eqnarray}\label{dashedmass}
\begin{array}{c}
m_{0}'=\lim \frac{\partial}{\partial~r}\left(m(\nu,r)\right) \\
\begin{tiny}\nu\rightarrow 0\end{tiny}\\
\begin{tiny}r\rightarrow 0\end{tiny}
\end{array}
\end{eqnarray}
and $\dot{m_{0}}$ is given by the eqn.(\ref{massd}).

Research by Dwivedi and Joshi in Refs.\cite{joshi2, joshi5} showed that, according to Tipler's definition, any classical singularity in Vaidya spacetime in the background of Einstein gravity should be an extremely severe curvature singularity. Furthermore, the authors have demonstrated that it is not always the case that the strong curvature singularities are trapped by horizons, as suggested by hypothesis \cite{tip2}. It is expected that we will get some deviations from the classical results for our BEC model. For our model, the mass parameter was calculated as
\begin{equation}\label{mmm1}
m(\nu,r)=\frac{a\alpha\sin{kr}}{k^3}+\frac{\beta r^3}{6}+\frac{\Lambda r^{3}}{6}+rh(\nu)+g(\nu)+\sigma(\nu) 
\end{equation}
We take the similar ansatz that we considered in the study of collapse, i.e., $h(\nu)=h_{0}\nu$, $g(\nu)=g_{0}\nu$, and $\sigma(\nu)=\sigma_{0}\nu$. Taking the derivative with respect to $\nu$ we get
\begin{equation}
\dot{m}(\nu,r)=rh_{0}+g_{0}+\sigma_{0}
\end{equation}
Now to get $\dot{m_{0}}$ we take the limit of $\dot{m}(\nu,r)$ as $\nu\rightarrow 0$ and $r\rightarrow 0$.  So we get,
\begin{eqnarray}\label{massd}
\begin{array}{c}
\dot{m_{0}}=\lim~ \dot{m}(\nu,r)=g_{0}+\sigma_{0} \\
\begin{tiny}\nu\rightarrow 0\end{tiny}\\
\begin{tiny}r\rightarrow 0\end{tiny}
\end{array}
\end{eqnarray}
Using the above expression in Eqn.(\ref{maha}) we get
\begin{equation}
S=\frac{1}{2}\chi_{0}^{2}\left(g_{0}+\sigma_{0}\right)  
\end{equation}
Here we see that the signature of $\left(g_{0}+\sigma_{0}\right)$ determines whether the singularity will be strong or weak. For $\left(g_{0}+\sigma_{0}\right)>0$ we have $S>0$ and the singularity formed is a strong singularity. Contrarily if $\left(g_{0}+\sigma_{0}\right)\leq 0$, then $S\leq 0$ and hence the singularity is a weak singularity. So for the BEC DM model, there is provision for both strong and weak singularities depending on the initial conditions. Here $\sigma_{0}$ comes from the contribution of Vaidya null radiation. Theoretically, there can be three cases. If we neglect the contribution from the null radiation then we have $\sigma_{0}=0$, and $S=\frac{1}{2}\chi_{0}^{2}g_{0}$. In this case, we get a strong singularity if $g_{0}>0$. When we have a positive contribution coming from the null radiation then $\sigma_{0}>0$. In this case, a strong singularity occurs for $g_{0}>0$ or $g_{0}=0$ or $g_{0}<0$ but $|g_{0}|<|\sigma_{0}|$. Finally, if there is a negative contribution coming from the null radiation, then we have $\sigma_{0}<0$. In this case, we get a strong singularity if $g_{0}>0$ and $|g_{0}|>|\sigma_{0}|$. Since the $g_{0}$ term depends on the initial conditions of the stellar evolution, probably the inherent properties of Bose-Einstein condensate determine the strength of the singularity. What is evident from the study is that in the theoretical framework, there is a possibility of both strong and weak singularities.

The causal behavior of the trapped surfaces evolving in spacetime during the collapse evolution typically determines the emergence of a naked singularity or black hole. The edge of the confined surface region in spacetime is known as the apparent horizon. For the generalized Vaidya spacetime, the equation of the apparent horizon is given by,
\begin{equation}
\frac{2m(\nu,r)}{r}=1
\end{equation}
Then the slope of the apparent horizon at the central singularity $(\nu\rightarrow 0, r\rightarrow 0)$ can be calculated as \cite{mken1}
\begin{equation}
\left(\frac{d\nu}{dr}\right)_{AH}=\frac{1-2m'_{0}}{2\dot{m}_{0}}
\end{equation}
Using the mass parameter given in Eqn.(\ref{mmm1}) we will get the slope of the apparent horizon for our model.

\section{Junction Conditions}
We consider a time-like 3D-hypersurface $\Sigma$ which divides the 4D spacetime into two distinct 4D manifolds $\mathcal{V}^{-}$ and $\mathcal{V}^{+}$ each of class $C^{4}$. Each of these manifolds $\mathcal{V}^{-}$ and $\mathcal{V}^{+}$ contain $\Sigma$ as its boundary. Let the coordinates in the regions $\mathcal{V}^{-}$ and $\mathcal{V}^{+}$ be given by $\chi_{-}$ and $\chi_{+}$ respectively. Moreover, the intrinsic coordinate on the boundary $\Sigma$ is given by $\xi^{a}$. Let the intrinsic metric to $\Sigma$ be $g_{ij}$ and the metric in $\mathcal{V}^{\pm}$ be $g_{\alpha\beta}^{\pm}$. The metric inside the collapsing matter which occupies the region $\mathcal{V}^{-}$ is given by a Friedmann-Robertson-Walker spacetime
\begin{equation}\label{metint}
ds_{-}^2=g_{\alpha\beta}^{-}d\chi^{\alpha}_{-}d\chi^{\beta}_{-}=-d\tau^{2}+a^{2}(\tau)\left[dr^{2}+r^{2}\left(d\theta^{2}+\sin{\theta}^{2}d\phi^{2}\right)\right]
\end{equation}
where $\tau$ is the proper time on co-moving world lines (along which $r$, $\theta$, $\phi$ are constants) and $a(\tau)$ is the scale factor. The metric outside the collapsing matter (which is the region $\mathcal{V}^{+}$) is given by the imploding generalized Vaidya metric
\begin{equation}\label{metext}
ds_{+}^{2}=g_{\alpha\beta}^{+}d\chi^{\alpha}_{+}d\chi^{\beta}_{+}=-\left[1-\frac{2m(\nu,r)}{r}\right]d\nu^{2}+2 d\nu dr+r^{2}\left(d\theta^{2}+sin^{2}\theta d\phi^{2}\right)
\end{equation}
where $m(\nu,r)$ is the gravitational mass of the collapsing star. As seen from the outside $\Sigma$ can be described by the parametric equations $\nu=T(\tau)$ and $r=R(\tau)$, where $\tau$ is the proper time with observers co-moving with the surface. This is the same $\tau$ that appears in the metric given in Eqn.({\ref{metint}}). We choose the coordinates $\xi^{a}=(\tau, \theta, \phi)$ on the hypersurface $\Sigma$. Consequently $e_{\tau}^{\alpha}=u^{\alpha}$ is the four-velocity of an observer co-moving with the surface of the collapsing star. 

Now Israel's junction conditions as described by Santos can be given by \cite{junction0, junction1, junction2}\\

(I) Continuity of the line element over the boundary, i.e.
\begin{equation}\label{condition1}
\left(ds_{-}^2\right)_{\Sigma}=\left(ds_{+}^2\right)_{\Sigma}=\left(ds^2\right)_{\Sigma}
\end{equation}

where $(~)_{\Sigma}$ means the value of $(~)$ on the boundary $\Sigma$.\\

(II) The continuity of the extrinsic curvature over the boundary surface $\Sigma$, i.e.,
\begin{equation}\label{condition2}
\left[K_{ij}\right]=K_{ij}^{+}-K_{ij}^{-}=0
\end{equation}

As seen from $\mathcal{V}_{-}$ the induced metric on the boundary $\Sigma$ is given by,
\begin{equation}
ds^{2}_{\Sigma}=g_{ij}d\xi^{i}d\xi^{j}=-d\tau^{2}+a^{2}(\tau)R^{2}(\tau)\left(d\theta^{2}+\sin{\theta}^{2}d\phi^{2}\right)
\end{equation}
Putting $A(\tau)=a(\tau)R(\tau)$ in the above metric we get,
\begin{equation}\label{metbound}
ds^{2}_{\Sigma}=-d\tau^{2}+A^{2}(\tau)\left(d\theta^{2}+\sin{\theta}^{2}d\phi^{2}\right)
\end{equation}
As seen from $\mathcal{V}_{+}$ the induced metric on the boundary $\Sigma$ is given by,
\begin{equation}
ds^{2}_{\Sigma}=-\left(F\dot{T}^{2}-F^{-1}\dot{R}^{2}\right)d\tau^{2}+R^{2}(\tau)\left(d\theta^{2}+\sin{\theta}^{2}d\phi^{2}\right)
\end{equation}
where $F=\left(1-\frac{2m(T,R)}{R}\right)$. According to Israel's junction conditions, the induced metric must be the same on both sides of the hypersurface. So we have, 
\begin{equation}
R(\tau)=A(\tau), ~~~~~~~~~~~~ F\dot{T}^{2}-F^{-1}\dot{R}^{2}=1
\end{equation}
$R(\tau)$ can be obtained from the first condition and the second condition can be solved for $\dot{T}$. From the second equation, we get
\begin{equation}
F\dot{T}=\sqrt{\dot{R}^{2}+F}=g(R,\dot{R},T)
\end{equation}
This equation can be integrated for $T(\tau)$ and the motion of the boundary in $\mathcal{V}^{+}$ is completely determined.

We can obtain the unit normal to $\Sigma$ from the relations $n_{\alpha}u^{\alpha}=0$, $n_{\alpha}n^{\alpha}=1$. From $\mathcal{V}^{-}$ we have, $u^{\alpha}_{-}\partial_{\alpha}=\partial_{\tau}$ and $n_{\alpha}^{-}dx^{\alpha}=a d\chi$. We have chosen $n^{\chi}>0$ so that $n^{\alpha}$ is directed towards $\mathcal{V}^{+}$. From $\mathcal{V}^{+}$, we have, $u^{\alpha}_{+}\partial_{\alpha}=\dot{T}\partial_{\nu}+\dot{R}\partial_{r}$ and $n^{+}_{\alpha}dx^{\alpha}=-\dot{R}d\nu+\dot{T}dr$, with a consistent choice for the sign.

The surface $\Sigma$ is the boundary of the interior distribution of matter and in this case, is given by,
\begin{equation}
f(r,\tau)=r-r_{\Sigma}=0
\end{equation}
where $r_{\Sigma}$ is a constant. Then the vector orthogonal to $\Sigma$ is $\frac{\partial f}{\partial \chi^{a}_{-}}$ and is given by,
\begin{equation}
\frac{\partial f}{\partial \chi^{a}_{-}}=\left(0,1,0,0\right)
\end{equation}
Using the first junction condition (\ref{condition1}) and the metrics (\ref{metint}) and (\ref{metbound}) we get
\begin{equation}
A(\tau)=r_{\Sigma}a(\tau)   
\end{equation}
The unit vector normal to the surface $\Sigma$ becomes
\begin{equation}\label{normal}
n_{a}^{-}=\left[0,a^{2}(\tau),0,0\right]
\end{equation}
The extrinsic curvature of the manifolds $\mathcal{V}^{\pm}$ is given by
\begin{equation}\label{curvature}
K_{ij}^{\pm}=-n_{a}^{\pm}\frac{\partial^{2}\chi^{a}_{\pm}}{\partial \xi_{\pm}^{i}\partial \xi_{\pm}^{j}}-n_{a}^{\pm}\Gamma^{a}_{~bc}\frac{\partial\chi^{b}_{\pm}}{\partial \xi_{\pm}^{i}}\frac{\partial\chi^{c}_{\pm}}{\partial \xi_{\pm}^{j}}
\end{equation}
The extrinsic curvature component $K_{ij}^{-}$ of the boundary $\Sigma$ can be obtained using (\ref{metint}), (\ref{normal}) and (\ref{curvature}) as,
\begin{equation}\label{ext1}
K_{\tau\tau}^{-}=0
\end{equation}

\begin{equation}\label{ext2}
K_{\theta\theta}^{-}=\left(ra^{2}(\tau)\right)_{\Sigma}
\end{equation}

\begin{equation}\label{ext3}
K_{\phi\phi}^{-}=K_{\theta\theta}^{-} \sin^{2}{\theta}
\end{equation}
The above quantities are valid on the surface $\Sigma$.

For the exterior spacetime $\mathcal{V}^{+}$, the equation of the surface $\Sigma$ is given by,
\begin{equation}
f(\nu,r)=r-r_{\Sigma}(\nu)=0
\end{equation}
So the vector orthogonal to $\Sigma$ is given by
\begin{equation}
\frac{\partial f}{\partial \chi^{a}_{+}}=\left(-\frac{dr_{\Sigma}}{d\nu},1,0,0\right)
\end{equation}
Then the unit normal vector on the boundary $\Sigma$ is given by
\begin{equation}\label{normalext}
n_{a}^{+}=\left(1-\frac{2m}{r_{\Sigma}}-2\frac{dr_{\Sigma}}{d\nu}\right)^{-1/2}\left(-\frac{dr_{\Sigma}}{d\nu},1,0,0\right)
\end{equation}
Using the first junction condition (\ref{condition1}) and the metrics (\ref{metext}) and (\ref{metbound}) we can write the junction conditions in a different way as,
\begin{equation}
r_{\Sigma}(\nu)=A(\tau)
\end{equation}
and
\begin{equation}\label{jj1}
\left(1-\frac{2m}{r_{\Sigma}}-2\frac{dr_{\Sigma}}{d\nu}\right)_{\Sigma}=\left(\frac{1}{\dot{\nu}^{2}}\right)_{\Sigma}
\end{equation}
where $\dot{\nu}$ represents $\frac{d\nu}{d\tau}$.
Using the above equation in the expression for the normal in Eqn.(\ref{normalext}) we get the simplified form as,
\begin{equation}\label{normall}
n_{a}^{+}=\left(-\dot{r},\dot{\nu},0,0\right)
\end{equation}
Now using the metric (\ref{metext}) and the Eqns.(\ref{curvature}) and (\ref{normall}) we get the non-vanishing components of the extrinsic curvature as
\begin{equation}\label{ext4}
K_{\tau\tau}^{+}=\left(\frac{\ddot{\nu}}{\dot{\nu}}-\dot{\nu}\frac{m}{r^{2}}+\dot{\nu}\frac{m_{r}}{r}\right)_{\Sigma}
\end{equation}

\begin{equation}\label{ext5}
K_{\theta\theta}^{+}=\left(\dot{\nu}\left(1-\frac{2m}{r}\right)-r\dot{r}\right)_{\Sigma}
\end{equation}

\begin{equation}\label{ext6}
K_{\phi\phi}^{+}=\sin^{2}{\theta}K_{\theta\theta}^{+}
\end{equation}
the above expressions are valid on the surface $\Sigma$.

Equating the appropriate extrinsic curvature components for the manifolds $\mathcal{V}^{\pm}$ using the Eqns.(\ref{ext1})-(\ref{ext3}) and Eqns.(\ref{ext4})-(\ref{ext6}) we get
\begin{equation}
\left(\frac{\ddot{\nu}}{\dot{\nu}}-\dot{\nu}\frac{m}{r^{2}}+\dot{\nu}\frac{m_{r}}{r}\right)_{\Sigma}=0
\end{equation}

\begin{equation}
\left(\dot{\nu}\left(1-\frac{2m}{r}\right)-r\dot{r}\right)_{\Sigma}=\left(ra^{2}(\tau)\right)_{\Sigma}
\end{equation}

The variable $\tau$ can be removed from the equations above because it was defined as an intermediate variable. Consequently, the necessary and sufficient conditions on the spacetimes for the validity of the first junction condition is given by
\begin{equation}
\left(\frac{\ddot{\nu}}{\dot{\nu}}-\dot{\nu}\frac{m}{r^{2}}+\dot{\nu}\frac{m_{r}}{r}\right)_{\Sigma}=0
\end{equation}

\begin{equation}\label{j2}
\left(\dot{\nu}\left(1-\frac{2m}{r}\right)-r\dot{r}\right)_{\Sigma}=\frac{r_{\Sigma}^{2}(\nu)}{(r)_{\Sigma}}
\end{equation}
We obtain an expression for $m(\nu,r)$ from the above equation (\ref{j2}) as
\begin{equation}
m(\nu,r)|_{\Sigma}=\left[\frac{r}{2}\left\{1-\left(\frac{r^{2}(\nu)}{r}+r\dot{r}\right)\frac{1}{\dot{\nu}}\right\}\right]_{\Sigma}   
\end{equation}
We can eliminate $\dot{\nu}$ using Eqn.(\ref{jj1}) and get
\begin{equation}
m(\nu,r)|_{\Sigma}=\left[\frac{r}{2}\left\{1-\left(\frac{r^{2}(\nu)}{r}+r\dot{r}\right)\left(1-\frac{2m}{r}-2\frac{dr(\nu)}{d\nu}\right)^{1/2}\right\}\right]_{\Sigma}   
\end{equation}
If we consider $\dot{r}\approx 0$ on the surface $\Sigma$, then the above expression becomes
\begin{equation}
m(\nu,r)|_{\Sigma}=\left[\frac{r}{2}\left\{1-\frac{r^{2}(\nu)}{r}\left(1-\frac{2m}{r}-2\frac{dr(\nu)}{d\nu}\right)^{1/2}\right\}\right]_{\Sigma}   
\end{equation}
The above expression can be interpreted as the total gravitational mass of the star within the surface $\Sigma$. Intuitively this mass should match with the mass calculated from the field equations in Eqn.(\ref{sol2}) at the boundary $\Sigma$. So we have the following relation
\begin{equation}
\left[\frac{a\alpha\sin{kr}}{k^3}+\frac{\beta r^3}{6}+\frac{\Lambda r^{3}}{6}+rh(\nu)+g(\nu)+\sigma(\nu)\right]_{\Sigma}=\left[\frac{r}{2}\left\{1-\frac{r^{2}(\nu)}{r}\left(1-\frac{2m}{r}-2\frac{dr(\nu)}{d\nu}\right)^{1/2}\right\}\right]_{\Sigma}
\end{equation}
Using $h(\nu)=h_{0}\nu$, $g(\nu)=g_{0}\nu$ and $\sigma(\nu)=\sigma_{0}\nu$ we have from the above equation
\begin{equation}
r^{4}(\nu)\left(1-\frac{2m}{r_{\Sigma}}-\frac{2dr(\nu)}{d\nu}\right)=\left[1-\frac{2}{r_{\Sigma}}\left(\frac{a\alpha\sin{kr_{\Sigma}}}{k^3}+\frac{\beta r_{\Sigma}^3}{6}+\frac{\Lambda r_{\Sigma}^{3}}{6}+r_{\Sigma}h_{0}\nu+g_{0}\nu+\sigma_{0}\nu\right)\right]^{2}r_{\Sigma}^{2}
\end{equation}
If $m$ is considered to be of constant value on the surface $\Sigma$ then the above equation becomes a differential equation of $r(\nu)$ with respect to $\nu$, which may be solved to get $r(\nu)$.

\begin{figure}[hbt!]
\begin{center}
\includegraphics[height=6in]{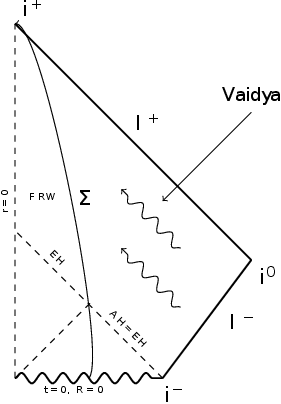}
\caption{Penrose diagram for matching the Vaidya and the FRW spacetimes. In the figure AH and EH stand for the apparent and the event horizons respectively. $\mathcal{I}^{\pm}$ represent the future (+) and the past (-) null infinity. $i^{\pm}$ represents the future (+) and past (-) timelike infinity. Moreover, $i^{0}$ represents the spacelike infinity. $\Sigma$ represents the boundary surface between the two spacetimes or the surface of the star. The curvy line at $t=0, R=0$ represents the singularity.}
\label{figepenrose}
\end{center}
\end{figure}

In Fig.(\ref{figepenrose}) we have shown the Penrose diagram for the matching of the Vaidya and the FRW spacetimes to gain more insight into the collapsing system. It represents graphically the causal structure of the spacetimes allowing us to visualize the geometry and the paths of the light rays and particles as they move through the spacetimes. The figure also helps in understanding and analyzing the properties of spacetime, especially the existence of singularities and the behavior of the light rays. In this case, it also helps in the comparison of the two different spacetimes. In the diagram, we see that the imploding Vaidya null radiation is depicted with curvy lines moving from the Vaidya spacetime region towards the boundary $\Sigma$. This null radiation is accompanied by the Bose-Einstein condensate considered in our model which moves towards the boundary forming a collapsing scenario. The center of the star is initially represented by the vertical timelike curve $r=0$, which is actually a spacetime point. After the singularity is formed at $t=0, R=0$ (shown by the curvy line) the center of the star is represented by a spacelike horizontal curve which represents a slice of time and not an actual spacetime point. The event horizon is first formed inside the boundary $\Sigma$ and as the singularity settles down the apparent horizon moves towards the event horizon before coinciding with it represented by $AH=EH$.

\section{Conclusion and Discussion}
In this work, we have explored a gravitational collapse in generalized Vaidya spacetime filled with Bose-Einstein dark matter. Investigating the prospect of a collapse without singularity is the main goal. We sought to explore if trapped surfaces may form as a result of collapse (in the event of the formation of a singularity). We used the Vaidya metric to represent the spacetime of a massive star. This choice was taken because the time-dependent Vaidya metric accurately describes the spacetime of a real star. Furthermore, because a collapsing process is a time-dependent event, it fits this metric perfectly. After computing the field equations, the mass term was calculated. Using this mass, we developed a theoretical framework for the mechanism of collapse.  The intention was to investigate the possibility of radial null geodesics extending outward from the core singularity. We arrived at a differential equation in $\nu$ and $r$ during the process. We took into consideration a limiting condition as $\nu\rightarrow 0$ and $r\rightarrow 0$ in order to investigate the effects close to the singularity. The type of singularity that emerged was actually governed by the signature of the roots of an algebraic equation in terms of a collapsing parameter $\chi_{0}$. When $\chi_{0}$ was positive, it meant that there were outgoing radial null geodesics, which ruled out trapped surfaces, or black holes. This would suggest that either no singularity was developed or, if it has, it is a naked singularity. The lack of any outgoing null geodesics was indicated if $\chi_0$ had negative or imaginary values. This implied the likelihood of an event horizon forming, implying the presence of a black hole.

For our Bose-Einstein condensate DM model it was found that the equation of state parameter $\alpha$ played a crucial role in determining the nature of singularity formed. It was seen that for $\alpha>0$ the collapse ends in a black hole, whereas for $\alpha<0$, the collapse results in the formation of a singularity that is globally naked. It was also seen that these effects were more prominent around the $k=0$ region where the roots acquired their maximum/minimum values. As $k\rightarrow \pm \infty$ the intensity of these effects slowly declines. So $k=0$ must be a very crucial point for the Bose-Einstein condensate model. A transition in the direction of the trajectories may correspond to a phase transition inside the dark matter. This is really an important piece of information regarding the model, considering that very little is known about the properties of dark matter due to their invisible nature. To gain more insight into the collapsing spacetimes we have dedicated a detailed section in the study of junction conditions where the exterior Vaidya metric is matched with a FRW interior metric. The junction conditions have been derived by choosing suitable coordinates for each region and we have gained valuable information from this study regarding the collapsing system. A Penrose diagram depicting the causal relations between the interior and the exterior spacetimes of the star has been shown.

Unlike black holes, naked singularities would not have an event horizon to hide them from view (if they existed). It is a challenging task to find visual clues to distinguish between the two because it goes beyond what is currently understood about physics and observational abilities. Nonetheless, there may be certain visible markers that help distinguish between black holes and naked singularities. The simplest method of differentiating between naked singularities and black holes is to determine if the event horizon is present or absent. The accretion disks encircling black holes and bare singularities may have different properties. For example, in the case of naked singularities, the absence of an event horizon may have an impact on the structure and features of the accretion disk. Different gravitational waves may be produced by a system with a bare singularity compared to one containing a black hole. Gravitational wave observations could reveal these changes and provide useful information. It is possible that a naked singularity doesn't attract matter around it as strongly as a black hole. Observing how nearby things move and the surrounding environment could reveal intriguing patterns. Bare singularities might not release Hawking radiation, in contrast to black holes. It could be possible to learn more about the existence of the postulated singularity by examining the radiation properties in further detail. A closer look at quantum events around the singularity could aid in differentiating between black holes and naked singularities. This would require a deeper understanding of quantum gravity, which is currently beyond the reach of scientists. Events around naked singularities could be more chaotic or exhibit unexpected behaviors if they violate the Cosmic Censorship Hypothesis, in contrast to events around black holes. It is important to note that the suggestions made above are theoretical in nature and that a strong theoretical framework for naked singularities is still lacking. Moreover, it is plausible that our current observational capacities may prove insufficient in locating and characterizing these anomalies. Such extreme astrophysical phenomena necessitate a deeper comprehension of the interplay between general relativity and quantum mechanics, a field of current theoretical physics study. Nevertheless, this work intends to improve our understanding along similar avenues. Moreover, the presence of Bose-Einstein condensate dark matter gives a new dimension to the study. Finally, from the standpoint of the cosmic censorship theory, this work appears to represent a significant advancement, providing a significant counterexample.

\section*{Acknowledgments}
The author acknowledges the Inter-University Centre for Astronomy and Astrophysics (IUCAA), Pune, India for granting visiting associateship.

\section*{Data Availability Statement}

No data was generated or analyzed in this study.

\section*{Conflict of Interest}

There are no conflicts of interest.

\section*{Funding Statement}

There is no funding to report for this article.


\end{document}